\documentclass[12pt,preprint]{aastex}

\def \arcsec{\hbox{$^{\prime\prime}$}}

\def \c2{cm$^{-2}$}

\def \kms{kms$^{-1}$}

\def \nh3{NH$_3$}
\def \n2h{N$_2$H$^+$}

\def \nh2{n_{H_2}}
\def \nh1{n_{HI}}

\def \h2{H$_2$}

\def \Ms{$M_{\odot}$}
\def \Ls{$L_{\odot}$}
\def \mic{$\mu$m}

\def \be{\begin{equation}}
\def \ee{\end{equation}}
\def \bf{\begin{figure}}
\def \ef{\end{figure}}

\begin{document}
\shorttitle{}
\shortauthors{Goldsmith, Pandian, \& Deshpande}

\title{A Search for 6.7 GHz Methanol Masers in M33 }

\author{Paul F. Goldsmith \altaffilmark{1}, Jagadheep D. Pandian \altaffilmark{2}, and 
Avinash A. Deshpande \altaffilmark{3}}
\altaffiltext{1}{Jet Propulsion Laboratory, California Institute of Technology, Pasadena CA; Paul.F.Goldsmith@jpl.nasa.gov}
\altaffiltext{2}{Department of Astronomy, Cornell University, Ithaca NY and Max-Planck-Institut f\"{u}r Radioastronomie, Aug dem H\"{u}gel 69, 53121, Bonn; jpandian@mpifr-bonn.mpg.de}
\altaffiltext{3}{Raman Research Institute, Bangalore; desh@rri.res.in }


\begin{abstract}

We report the negative results from a search for 6.7 GHz methanol masers in the nearby spiral galaxy M33.  
We observed 14 GMCs in the central 4 kpc of the Galaxy, and found 3$\sigma$ upper limits to the flux density of $\sim$9 mJy in spectral channels having a velocity width of 0.069 \kms.
By velocity shifting and combining the spectra from the positions observed, we obtain an effective 3$\sigma$ upper limit on the average emission of $\sim$1mJy in a 0.25 \kms\ channel.  
These limits lie significantly below what we would expect based on our estimates of the methanol maser luminosity function in the Milky Way.  
The most likely explanation for the absence of detectable methanol masers appears to be the metallicity of M33, which is modestly less than that of the Milky Way.  
\end{abstract}

\setcounter{footnote}{0}

\section{Introduction}

6.7 GHz methanol masers are the second brightest maser transition ever observed, and are typically much brighter than OH masers. 
The properties of methanol masers in the Magellanic clouds \citep{sinclair1992,ellingsen1994b,beasley1996} are consistent with those of our Galaxy, given appropriate consideration for different galactic properties such as metallicity.
The SMC has an oxygen abundance 12 + log(O/H) = 7.96 \citep{vermeij2002}, which is a factor $\simeq$ 6 smaller than that of GMCs in the Milky Way \citep{peimbert1993}.  

However, 6.7 GHz methanol masers have not been discovered in galaxies beyond the Magellanic clouds. 
In particular, surveys have shown that there is no analog to OH megamasers at 6.7 GHz 
\citep{ellingsen1994a,phillips1998,darling2003}. 
It would be of interest to detect methanol masers in a Milky Way like galaxy. 
If the number of sources detected were large, one could derive the methanol maser luminosity function since all sources would be at the same distance. 
Further, one could look for correlations of methanol masers with giant molecular cloud masses, other types of masers, etc. 
The Arecibo radio telescope offers unequaled sensitivity for targeted surveys for methanol masers, and the
nearest spiral galaxy which can be observed with this instrument is M33.

\begin{figure}
\begin{center}
\includegraphics[angle=0]{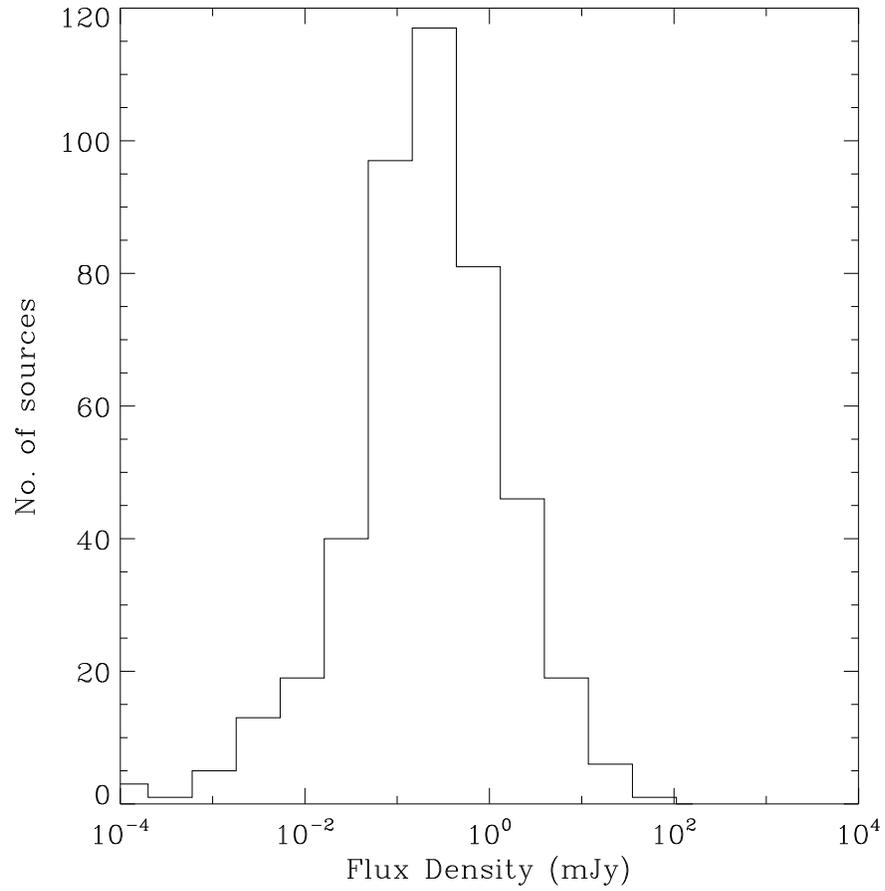}
\caption{Distribution of expected flux densities of methanol masers in M33 based on observed masers
in the Milky Way.}
\label{expected_hist}
\end{center}
\end{figure}

It is difficult to develop an optimal strategy to search for extragalactic methanol masers since the luminosity function of methanol masers in our Galaxy is unknown. 
This is mostly due to difficulties in determining distances to methanol masers, which is compounded by the kinematic distance ambiguity. 
We have adopted the following approach.  
We selected sources from the general catalog \citep{pestalozzi2005}, eliminating sources
wihtin 10 degrees of longitudes 0 and 180 degrees since these suffer from large uncertainties
in kinematic distance.  
Assuming that all masers are at their near kinematic distance, we then scaled the source flux density to the distance of M33, which we take to be 840 kpc, averaging the results of \cite{lee2002} and \cite{freedman1991}, which gives a result consistent with the distance determined by \cite{kim2002}. 
The resulting histogram of ``expected'' methanol maser flux densities is showin in Figure \ref{expected_hist}

We thereby obtained an estimate for the distribution of flux density of methanol masers in M33, which of course must be regarded with considerable skepticism, as the original Galactic sample has significant biases, not to mention the unknown differences in the properties of massive star formation in M33 and the Milky Way.
It is also inherently pessimisitic in the sense that some of the Galactic masers are at their far
kinematic distance, and hence are more luminous than we have assumed.
With these caveats, we found a ``high flux density tail'' for our hypothetical methanol maser population in M33 that extends from 100 mJy down to 1 mJy.  
Approximately 16\% of the total number of masers would be expected to have flux density greater than 1 mJy,
with about 6\% having flux density greater than 3 mJy.  


\section{Observations}

The search for methanol masers in M33 was carried out using the 305 m Arecibo telescope\footnote{The Arecibo Observatory is a facility of the National Astronomy and Ionosphere Center which is operated with support from the National Science Foundation}and the C-band high receiver \citep{pandian2006} between 2005 July 9 and July 21.  
The data were taken in standard Arecibo position switched mode in which the source is observed for five minutes, and then a reference position offset by 6$^m$ in Right Ascension is observed for an equal length of time.  
The spectrometer used was the interim correlator. 
Two boards recorded the two orthogonal linear polarization data at 0.069 \kms resolution (3.125 MHz bandwidth with 2048 lags) while the two other oboards recorded the same data at a higher velocity resolution of 0.034 \kms (1.5625 MHz bandwidth with 2048 lags).  
All the boards were centered on the CO velocity as given by \cite{engargiola2003}.  
The FWHM beam width of the telescope at 6.7 GHz is 40\arcsec, which corresponds to a linear size of 160 pc at a distance of 840 kpc.  
System temperatures varied between 23 K and 34 K, with approximately 20 percent difference between the two polarizations.  
The data were converted to flux density using the elevation dependent gain curve for this frequency, with a typical conversion factor of 5 K/Jy.  

Based on the correlation between methanol masers, massive star formation, and Giant Molecular Clouds GMC) in the Milky Way, the positions in M33 observed were the most massive clouds given in Table 1 of \cite{engargiola2003}. 
Each of the first 14 clouds, including all of the most massive GMCs in M33 (M $\geq$ 4$\times$10$^5$ \Ms), was observed for a total of 30 minutes on plus 30 minutes for the reference position.  
The resulting rms for the low resolution data after averaging the two linear polarizations is 3 mJy, and for the high resolution data is a factor of $\sqrt{2}$ higher. 
No spectral features having greater than 3$\sigma$ significance were observed.


\section{Data Analysis}

The observation of 14 massive GMCs in M33 has yielded no methanol maser detections at a 3$\sigma$ level of 9 mJy.  
Based on comparison with the expected distribution of flux densities based on 6.7 GHz methanol masers in the Milky Way, this is surprising.  
There are estimated to be approximately 1200 methanol masers in the Milky Way \citep{vanderwalt2005}, while the number of GMCs out to 8 kpc from the Galactic Center is 4000 to 5000 (Solomon \& Sanders 1980; Scoville \& Sanders 1987).  
Having observed 14 GMCs in M33 with mass between 2 and 7 $\times$10$^5$ solar masses, if the relationship between methanol masers and GMCs were the same as in the Milky Way, we would have expected to have made several detections.  

\begin{figure}
\begin{center}
\includegraphics[angle=270, width=\textwidth]{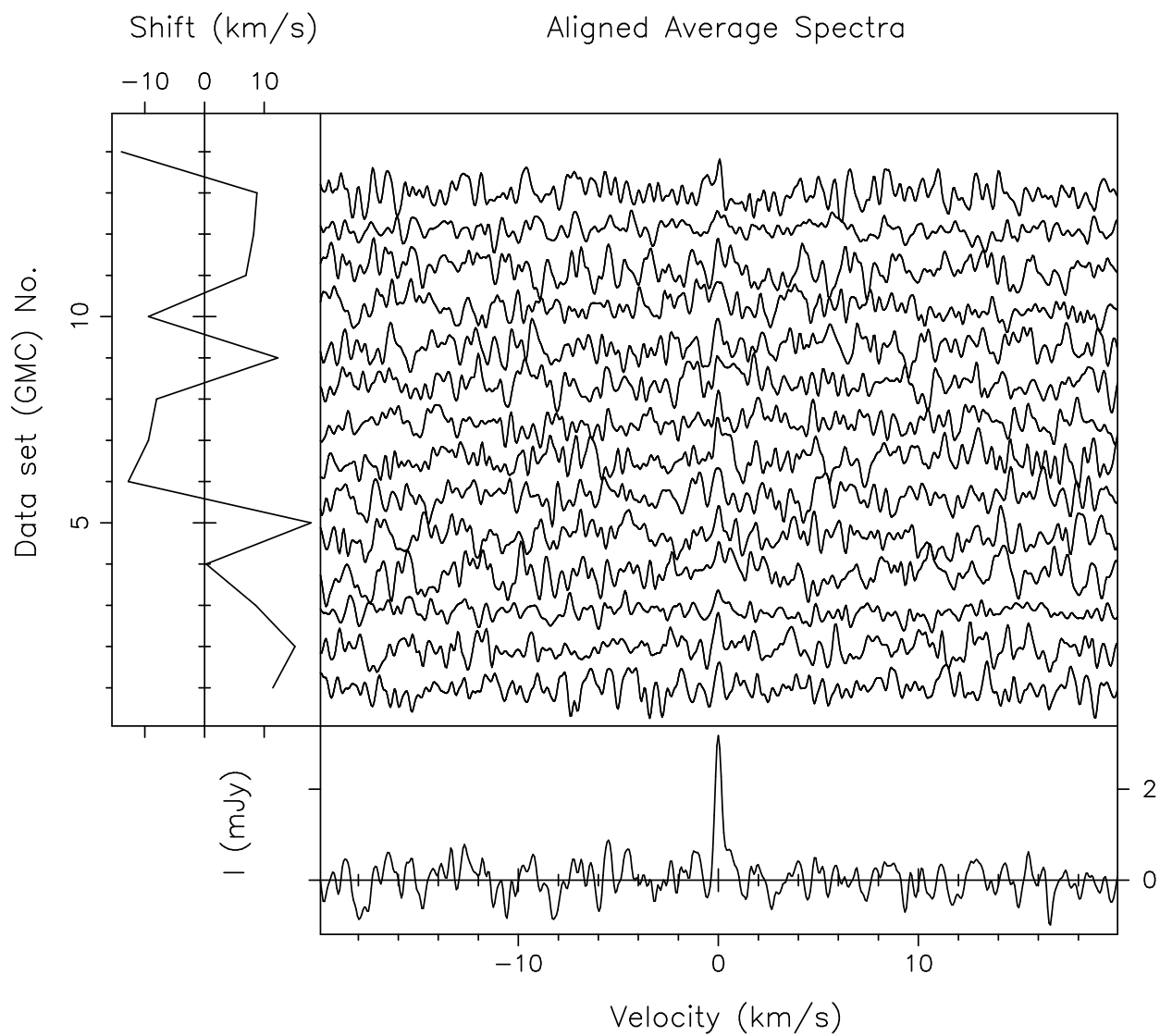}
\caption[The sum of aligned spectra for 14 GMCs in M33: Spectrum A.]{The bottom panel shows the result of adding the spectra of 14 GMCs that have been shifted in velocity to the velocity of peak cross-correlation with a Gaussian template. This spectrum is denoted spectrum A. The x-axis of the plot shows the velocity offset from that of the CO emission line. The top panel shows the shifted individual GMC spectra, each of which is smoothed to a linewidth of 0.25 \kms. The side panel shows the amount of velocity shift applied for each spectrum.}
\label{m33a}
\end{center}
\end{figure}

\begin{figure}
\begin{center}
\includegraphics[angle=270,width=\textwidth]{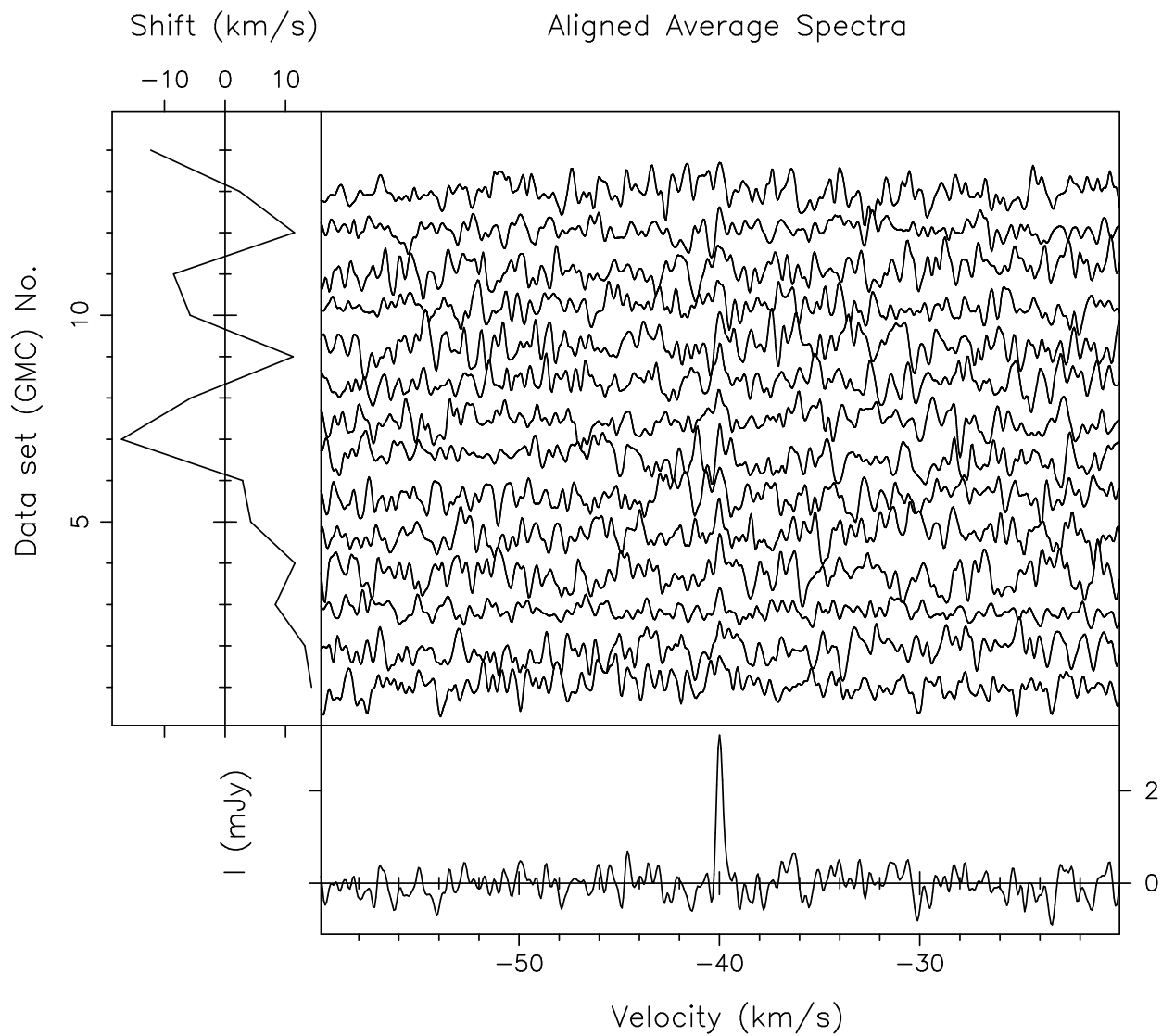}
\caption[The sum of aligned spectra for 14 GMCs in M33: Spectrum B.]{The bottom panel shows the result of adding the spectra of 14 GMCs that have been shifted in velocity to the velocity of peak cross-correlation with a Gaussian template. This analysis has been done over a portion of the spectrum where no methanol maser emission is expected and the result is denoted spectrum B. The x-axis of the plot shows the velocity offset from that of the CO emission line. The top panel shows the shifted individual GMC spectra, each of which is smoothed to a linewidth of 0.25 \kms. The side panel shows the amount of velocity shift applied for each spectrum.}
\label{m33b}
\end{center}
\end{figure}

Assuming that all the GMCs do exhibit methanol maser emission at a weak level, one can combine the data for all GMCs to get a better upper limit on the emission. 
This assumes that the emission occurs at the same velocity offset with respect to the velocity of the central channel in any given spectrum, which is not necessarily true. 
To allow for different velocity offsets for the location of emission in different GMCs, we follow the following procedure. 
We first select a section of the spectrum around the velocity of CO emission for each GMC ($\sim \pm 40$ km s$^{-1}$), which includes the largest offset one would expect between the methanol maser and the molecular line emission (expected to be less than 15 km s$^{-1}$). 
Each sub-spectrum is then cross-correlated with a Gaussian whose width represents the expected linewidth of the maser emission ($\sim 0.30$ \kms). 
Each sub-spectrum is then shifted by the offset between the central channel and the channel of peak cross-correlation. 
Following this procedure, the shifted data for all GMCs are added together. 
We denote the resulting spectrum as ``A'', and is shown in Figure \ref{m33a}.
The above procedure will create a line feature resembling the correlation template (a Gaussian signal in this case) from the constructive alignment of purely random noise. 
Hence, the procedure is repeated for a different portion of the spectrum, where one does not expect to see any methanol maser emission, and the resulting spectrum is denote ``B'' (Figure \ref{m33b}). 
We then compare the A and B spectra. 
A weak emission feature would be manifest by the peak in the spectrum A being higher than the peak in spectrum B by a statistically significant amount. 
Since this is not the case as can be seen by comparing Figures \ref{m33a} and \ref{m33b}, we conclude that there is no detectable methanol maser emission in the data. 
The resultant 3 $\sigma$ upper limit on the average emission is $\sim 1$ mJy. 
This is a full order of magnitude below the flux density that we would expect to be common in M33 from the extrapolation of our admittedly uncertain results in the Milky Way. 
This difference, if real, would certainly suggest some large-scale difference in the molecular clouds or massive young stars in M33 compared to the Milky Way. A much more meaningful result could be obtained if we had better knowledge of the methanol maser luminosity function in our Galaxy. 


\section{Discussion and Conclusions}

It is difficult to be highly quantiative about the lack of detection of methanol maser emission in M33 beyond the statements given above.  
Given that the 40\arcsec\ Arecibo beam subtends a region 160 pc in size at the distance of M33, we should be sensitive to a methanol maser located anywhere within the individual giant molecular clouds being observed.  
Based on our estimate of the Milky Way methanol maser luminosity function, we would expect a significant fraction of masers to have a flux density $\geq$ 10 mJy, significantly higher than the individual limits we have obtained, and an order of magnitude greater than the averaged limit derived by combining the results from the 14 GMCs observed.

Explanations of the rarity of methanol masers in external galaxies have focused on (1) insufficient methanol density over path where amplification could take place and (2) insufficient pumping to invert the methanol transition in question \citep{phillips1998}. 
Both of these can result from low metallicity.  
A reduced abundance of oxygen, for example, will likely reduce both the abuance of methanol (CH$_3$OH) and of dust, which is required to convert the short wavelength radiation from massive young stars to the infrared wavelengths required for maser pumping.  
The rarity of masers in the Magellanic clouds has been discussed in similar terms by \cite{beasley1996}.  
The O/H ratio in the Magellanic clouds is dramatically lower than that of the Milky Way (see discussion in Beasley et al. 1996 and the more recent measurement by Vermeij \& van der Hulst 2002).  
For M33, the O/H ratio is only slightly less than that of the Milky Way.
\cite{vilchez1988} determined that 12 + log(O/H) = 9.0 at the center of M33, falling to $\sim$8.5 at distances between 2 and 5 kpc from the center of the galaxy.
The best fit line of \cite{crockett2006} to their new data plus previous data has a very small slope of -0.12 dex kpc$^{-1}$ and 12 + log(O/H) = 8.3 at 5 kpc from the center of M33.
These values may be compared to Galactic values of 12 + log(O/H) = 8.51 for Orion and 8.78 for M17 \citep{peimbert1993}.
The GMCs we have observed are located between 0.5 and 4 kpc from the center of M33, with a mean distance of 1.9 kpc.  
We conclude that the relevant O/H ratio in M33 is between 0.1 and 0.4 dex below that of the Milky Way.  
This difference is much smaller than that between the SMC and the Milky Way, for which log(O/H) differs by $\simeq$ 0.8 dex.
If the relatively small difference between the metallicity in the Milky wand M33 is responsible for the lack of methanol masers in the latter, it suggests that the maser luminosity must be a very senstive function of the galactic metallicity.

\cite{henkel1987} have detected thermal methanol emission from two galaxies, IC342 and NGC253, finding that the fractional abundance of methanol is $\simeq$10$^{-7.5}$.  
This is similar to that found in GMCs in the Milky Way, but is a factor of at least 100 lower than that required for high brightness methanol maser luminosity as discussed by \cite{sobolev1997}.
However, this difference is also found in comparing hot cores in the Milky Way, presumed to be the sites of methanol masers, with more extended molecular cloud regions. 
The methanol abundance may be greatly increased in regions near hot stars by thermal desorption of molecules frozen onto grain surfaces. 
We have no direct evidence regarding the methanol abundance in M33, so it is possible that the lower O/H
ratio does yield an insufficient methanol abundance to produce highly luminous masers.

There are individual regions within M33, presumably powered by massive young stars \citep{hinz2004}, which are powerful far--infrared sources.  
The infrared flux is a critical requirement for maser pumping as elaborated in models of \cite{cragg1992}, \cite{sobolev1994}, and \cite{sobolev1997}.
The transition to the second torsional excited state occurs at a wavelength of $\simeq$ 30 \mic, so that dust temperatures of at least 150 K are required to achieve high maser brightness \citep{sobolev1997}.  
Among our targets was number 8 of \cite{engargiola2003}, which lies within 15\arcsec of the optical nebula NGC 604.
It has an IRAS luminosity of 6.8$\times$10$^7$ \Ls \citep{rice1990} and the associated GMC has a mass derived from CO of 4.4$\times$10$^5$ \Ms \citep{engargiola2003}.
By Galactic standards this region would seem likely to harbour a high luminosity methanol maser, but no such
emisson was detected.

\cite{fix1985} were unsuccessful in a search using the VLA for highly luminous OH masers in M33.
Their limit of $\simeq$ 25 mJy (5$\sigma$) was sufficient to elminate the presence of any type I maser
as luminous as the brightest type I masers in the Milky Way
The present work thus reemphasizes the mystery of the lack of luminous masers in M33.

\acknowledgments

We are grateful to Phil Perillat for many discussions regarding the performance of the Arecibo telescope, and routines for supplying routines for data reduction.  We thank German Cortes--Medellin and Lynn Baker for their contributions to the receiver system development, Kurt Kabelac and David Overbaugh for machining most of the dewar components, and Rajagopalan Ganesan and Lisa Locke for calibrating the C-band high receiver used in this work. We are also grateful to numerous other members of staff at the Arecibo Observatory who helped us in the receiver installation and calibration, and in setting up our observations. This work was supported in part by the Jet Propulsion Laboratory, California Institute of Technology. This research has made use of NASA's Astrophysics Data System.

\end{document}